\documentclass{aa}
\usepackage{graphicx}

\begin{document}

\title{\bf The origin of the radio emission from $\beta$~Lyrae}
\author{      G. Umana \inst{1} 
         \and F. Leone  \inst{2} 
         \and C. Trigilio \inst{1} 
	}
\institute{Istituto di Radioastronomia del C.N.R., Stazione VLBI di Noto, 
           C.P. 161 Noto, Italy
	\and 
          INAF,  Osservatorio Astrofisico di Catania, Via S. Sofia 78,
           95123 Catania, Italy
	}
\offprints{G. Umana, umana@ira.noto.cnr.it}
\mail{umana@ira.noto.cnr.it}
\date{Received \hbox to1in{\hrulefill}}
  \abstract{In this paper we  present new observational evidence
that supports the presence of an extra source of continuum emission in the 
binary system \object{$\beta$~Lyrae}.
New VLA and IRAM observations, together with published data from the literature and 
ISO archive data, allow us to build the Spectral Energy Distribution of the 
binary between $5\times 10^{9}$Hz and $5\times 10^{15}$Hz.
The radio-millimeter part of the spectrum is consistent with free-free 
emission from a symbiotic-like  wind associated with the primary component 
and ionized by the radiation field of the hidden companion. 
Furthermore, we also consider the possibility that the observed radio flux originates from collimated radio structures associated with the mass gaining component
and its disk (Conical thermal jets).
  An extrapolation of this emission to the far-IR part of the spectrum 
indicates that in both cases the free-free contribution at these frequencies cannot explain the observations and that the observed 
infrared excess flux is due principally to the secondary component and its 
associated disk.
}
     
\authorrunning{Umana et al.}
\maketitle

%
\section{Introduction}

Even if the eclipsing binary \object{$\beta$~Lyrae} is one of the most studied 
stellar systems,  its  enigmatic spectroscopic and photometric behaviour 
is not fully understood.
The  current view is a non-degenerate, semi-detached interacting binary system 
in the phase of large-scale mass transfer between components.
The mass losing component  is a  B6-B8p II, while the  unseen 
mass gaining component is probably a B0V star,  embedded in an 
optically thick accretion disk (Hubeny \& Plavec \cite{Hupla91}).
The presence of a large plasma cloud, surrounding both components,
has been inferred from optical and UV emission lines (Batten \& Sahade 
\cite{Batten}; Hack et al. \cite{Hack}) as well as from the analysis of 
UV light curves (Kondo et al. \cite{Kondo}). 
Recently  the presence of  jet-like structures, probably related to the 
accretion disk,  has been shown  by the studies of Harmanec et al. 
(\cite{Harmanec96}) and confirmed by Hoffman et al. (\cite{Hoff1998}). 
\object{$\beta$~Lyrae} is also a well known radio source. 
First detected, in the early seventies, by Wade \& Hjellming (\cite{Wade72}) 
at 2.7 and 8.1 GHz, it was then monitored, at two frequencies,  by Gibson 
(\cite{Gibson75}). 
The source always exhibited a thermal-like spectrum but Wright \& Barlow 
(\cite{Wri1975}) pointed  out that the observed slope of the radio spectrum 
($\alpha$=0.96) was intermediate between the slope expected for a simple 
\ion{H}{ii} region and from a stellar wind, implying that the physics underlying
the radio emission of \object{$\beta$~Lyrae} is more complicated than assumed 
in either of these models.

Recent high resolution MERLIN observations of \object{$\beta$~Lyrae} at 4.9 $GHz$ 
(Umana et al. \cite{Umana}) have revealed an extended radio nebula around the 
system, whose brightness temperature ($1.1 \times 10^{4}$K) confirms the 
thermal origin of the radio emission.
Such a nebula can be re-conducted to a massive wind associated with the cooler 
primary ionized by the hotter secondary if the two-winds model by Mazzali et 
al. (\cite{Mazza92}) is adopted.

To assess a clear picture of the radio 
properties of \object{$\beta$~Lyrae}, an analysis of its spectral energy 
distribution (SED) from radio to infrared  appears to be necessary.
In this paper we  present multi-frequency VLA  and IRAM observations
that will be combined with data  from the literature in order to investigate the 
origin of the radio emission of \object{$\beta$~Lyrae}.

In the following, assuming the most common notation, we will indicate as primary
the mass losing component, as it is the most luminous in the optical region.
\section{Observations and Results}
\subsection{The VLA data}
The observations were carried out  using the VLA\footnote{The Very Large
Array is a facility of the National Radio Astronomy Observatory which is 
operated by Associated Universities, Inc. under cooperative agreement with the
National Science Foundation} on May 5, 1996, from 07:58 to 12:46 UT.
We observed \object{$\beta$~Lyrae} at five frequencies, namely 4.8 ~(C-Band), 
8.4  ~(X-Band), 14.9 ~(U-Band) and  22.0 ~(K-Band)  and 43.0 (Q-band) GHz,
using two independent 50 MHz bands. 
The observations were performed in compact (D) configuration, splitting the 
array in two different subarrays: the first one included all the antennas (13) 
equipped with Q-band receivers; the other subarray was used  sequentially at 
each remaining  frequency. This ensures the necessary sensitivity for detection 
of weak sources.
The adopted antenna configuration provides a typical beam size of 
$\sim 12 \arcsec$, $\sim 7.5 \arcsec$, $\sim 4 \arcsec$, $\sim 2.5 \arcsec$
and $\sim 3 \arcsec$ at 4.8, 8.4, 14.9, 22 and 43 GHz respectively.

A typical observing cycle consisted of 15-20~min integration time, preceeded and 
followed by a 2-min observation of the phase calibrator. This basic sequence
was repeated at least 3 times in order to improve the signal to noise ratio.

For Q-band observations a slightly different observing strategy was followed.
Frequent checking for pointing was performed since systematic  errors
may be a significant part of the primary beam at 43~GHz. Moreover, in order to
minimise atmospheric effects on the phases, a much shorter observing cycle has been adopted. 
As phase calibrator, \object{1925+211} was chosen for the Q-band, while for the other frequencies we used \object{1850+284}. 
The flux density scale was determined by observing \object{3C286}. The 43~GHz
flux of 1925+211 was determined relative to \object{3C286} by using only 
those scans obtained at the same elevation as for \object{3C286}. This ensures 
a  careful
amplitude calibration since at high frequencies there is a strong dependence on elevation of the  antenna aperture efficiency and of the atmospheric opacity.  

The data  processing was performed using the standard programs 
of the NRAO {\bf A}stronomical {\bf I}mage  {\bf P}rocessing  {\bf S}ystem (AIPS).
To achieve the highest possible signal to noise ratio, the mapping process was performed by using the natural weighting and the dirty 
map was {\bf CLEAN}ed down as close as possible to the theoretical noise.
At each frequency, the source position and the flux density determination were 
obtained by fitting a gaussian brightness distribution (JMFIT). 
To estimate the noise level in the maps  we analyzed an area on the map, with dimensions of $\sim 15-20 ~\theta_\mathrm{syn}$, away from the phase center.
Its consistency with the expected theoretical noise was always acertained. 

We detected \object{$\beta$~Lyrae} at all the 5 frequencies as a compact, 
unresolved source. Our results are summarized in Table~\ref{vla}, where the 
radio flux density, with its associated rms {\bf and the time of the observation are } reported for each different 
frequency.
In spite of the fact that \object{$\beta$~Lyrae} was one of the first stars to 
be detected in the early seventies, the measurements here reported are the 
first detections of the system at high frequencies.

\begin{center}
\smallskip
\begin{table}
\caption{VLA Observations}
\label{vla}
\begin{tabular}{cccc} \hline 
 Band [GHz]  & Flux density [mJy]  & rms [mJy] & U.T. \\ \hline 
4.8          &    4.27             &   0.06  & 10:40  \\
8.4          &    6.50             &   0.05  & 11:00  \\
14.9         &    9.60             &   0.20  & 10:00  \\
22           &   12.70             &   0.30  & 11:20  \\
43           &   13.90             &   0.50  & 10:00  \\     
\hline
\end{tabular}
\end{table}

\end{center}
 
\subsection{The IRAM data}
  Observations of the millimeter continuum of \object{$\beta$~Lyrae} were 
obtained in 1996, between 15 and 16 of May, with the IRAM\footnote {IRAM is supported by INSU/CNRS (France), MPG (Germany) and IGN (Spain)} Plateau de Bure (PdB)
interferometer (Guilloteau et al. \cite{Guilloteau}). These observations were carried out in a compact configuration (D), providing a typical beam size of 
$5\arcsec \times 8\arcsec$ at 1.3~mm.
 Since each of the four operating antenna was  equipped with a SIS dual-frequency receivers at 1.3 and 3~mm, the observations  were performed  simultaneously at  85~GHz and 215~GHz (on  May 15) and at 115~GHz and 240~GHz (on  May 16).
The observations were made in double-sideband (DSB) mode, with each sideband
of 500 MHz separated by 3~GHz. 

A  typical observation consisted of 20 minutes
on  source scan,  plus  several calibration
measurements to assure an accurate calibration of the interferometer, bandpass
and phase
calibration and a high pointing accuracy.  This sequence was repeated several times, for a total of
about four hours  source integration time,  in order to improve the signal to noise ratio. 
Phase calibration was performed by using \object{2013+370} while the flux scale
was fixed by using daily observations of the  primary calibrators 
\object{MWC349} and \object{3C454.3}.
The IRAM data were reduced and mapped  following the standard procedures
using the software developed by the Observatoire de Grenoble and IRAM
({\bf G}renoble {\bf I}mage and  {\bf L}ine {\bf D}ata {\bf A}nalysis {\bf S}oftware) 

We detected \object{$\beta$~Lyrae} at all the 4 frequencies. The results are 
summarized in Table~\ref{iram}, where the millimetric flux density, with its 
associated rms, is reported for each different frequency.

\begin{center}
\smallskip
\begin{table}
\caption{IRAM-PdB Observations}
\label{iram}
\begin{tabular}{cccc} \hline 
 Band [GHz]  & Flux density [mJy]  & rms [mJy] & U.T. \\ \hline 
85           &    26.8             &   0.7     &  5:13\\
115          &    27.0             &   4.0     & 4:54 \\
215          &    36.9             &   4.5     &  5:13\\
240          &    38.0             &   3.0     &  4:54\\
\hline
\end{tabular}
\end{table}

\end{center}

\section{Spectral energy distribution}

\smallskip

\object{$\beta$~Lyrae} has been observed by the ISO satellite and we 
retrieved three scientifically validated  spectra obtained with the Short Wavelengths Spectrograph
(SWS) from November 1997 to March 1998.
The spectra were successively analyzed  using the ISAP package.
This software was  specifically written for the reduction and scientific analysis of ISO SWS and LWS
(Long Wavelengths Spectrograph) Auto Analysis Results (AARs), 
which is the final output of the standard ISO pipeline processing.

\begin{figure*}
\resizebox{\hsize}{!}{\includegraphics{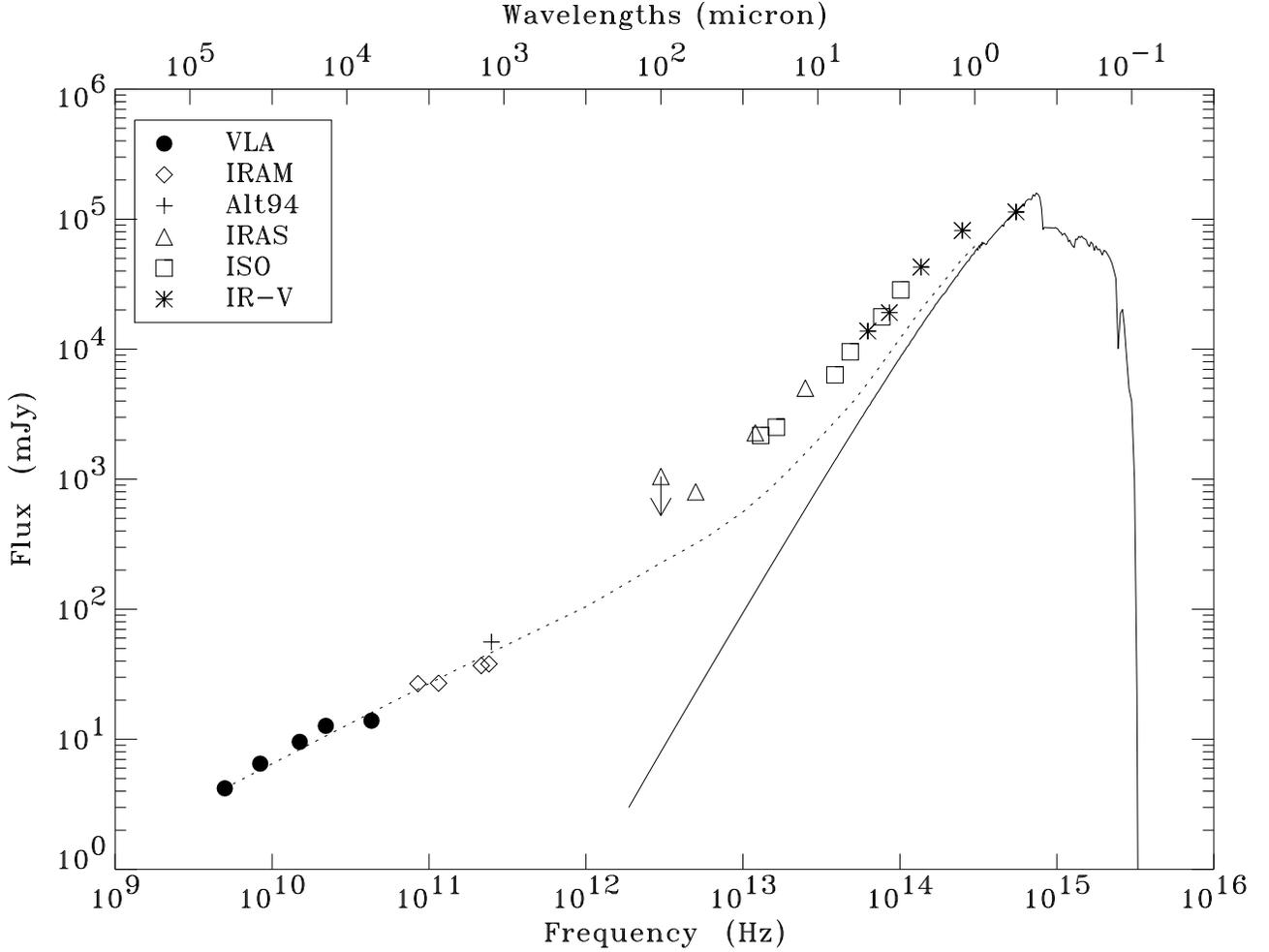}} 
\caption{
The SED of $\beta$~Lyrae. Dots, diamonds and the crosses represent VLA  and IRAM
 data from this paper plus a datum from Altenhoff et al. (\cite{Alte}).
Triangles and squares indicate IRAS and ISO data, while asterisks are VJKLM magnitudes from Viotti et al. (1978), V magnitude has been corrected as in the text. Data points have been corrected for interstellar extinction  following Mathis (\cite{Mathis90}) and assuming a $E_\mathrm{B-V}=0.^{m}065$.
The dotted line is the contribution at high frequencies of the free-free plus bound-free emission from the stellar wind.
The stellar contribution is represented 
by an ATLAS9 (Kurucz \cite{Kurucz93}) model, with  fluxes normalized to the flux in the V band (solid line).}
\label{sed}
\end{figure*}

As a final step, we estimate the continuum flux in 6 different channels by choosing areas free of  spectral lines and evaluating the average flux by 
applying a box-car filter of 2 $\mu$ width.
This analysis was performed only in the spectrum with the best signal to noise ratio.\\
\object{$\beta$~Lyrae} is also positionally associated with  the IRAS source 
\object{18482+3318} and, as pointed out by Friedmann et al. 
(\cite{Friedmann96}), has a spectral distribution in the IRAS bands typical of 
circumstellar material.\\
Fig.~\ref{sed} shows the {\bf S}pectral {\bf E}nergy {\bf D}istribution (SED) of \object{$\beta$~Lyrae} built  using  our reported observations, 
referenced measurements  from the literature 
as well as unpublished archive data.

\object{$\beta$~Lyrae} is reported in  the literature as a periodically variable source, whose ephemeris are (Harmanec \& Scholz \cite{Harmasch}):
\begin{equation}
T = JD ~2\,408\,247.966 + 12.91378 E + 3.87196 \times 10^{-6}E^{2}
\end{equation}
and  questions may arise regarding  whether it is possible to combine data obtained
at different orbital phases.

In order to
take in account  the variability with the orbital phase in building the SED,
we should consider only observations carried out at  orbital phases 
close to those of our VLA observations,  corresponding to orbital phase 
$\Phi \sim 0.151 \pm 0.016 $.

For the infrared and  visible, measurements  obtained at orbital phases between 0.14 and 0.16 and
corrected for an interstellar extinction of $E_\mathrm{B-V}=0.^{m}065$ 
(Abt et al., 1962)  have been used.

For the ISO and IRAS spectral regions, information on possible variation with orbital phase is
not available,  but  it is very probable that at longer wavelengths
the contribution of the circumsystem  material would be  important and
the resultant continuum will be less affected by  orbital variation.
Given the amplitude of light curves decreasing from the visible to the IR, proceeding  in a conservative way, if  we  use 
the eclipse depths reported by Zeilik et al. (\cite{Zeilik82}) 
for the infrared as dispersion  associated with each IRAS and ISO datum, 
we end up with an errorbar smaller than the symbols used in the plot of Fig.~\ref{sed}.

Finally, we can  built up the radio-mm spectrum quite confidentily, even if
the source has been reported variable at these frequencies and the data were 
not obtained  simultaneously,
since a VLA radio flux monitoring of \object{$\beta$~Lyrae} did not reveal any variations at least on a time-scale of a month or so (Umana et al. \cite{Umana2002}).
Moreover our millimetric measurements  are  in agreement with those obtained 
at 250~GHz by Altenhoff et al. (\cite{Alte}) more than 15 years ago.\\

Same secular changes of the radio flux density of \object{$\beta$~Lyrae} are present, as reported by several authors. However, in the following analysis of the spectral distribution we allow the flux density at 5~GHz to change between 
2.9 and 4.3 mJy, minimum and maximum measured  5~GHz flux density,  retaining the spectral index determined in the present work.
This leads to the conclusion that taking   secular changes into account does
not affect our main results.


\section{Discussion}

The radio  spectrum of \object{$\beta$~Lyrae} can be well 
represented by a single power-law.
The best fit of the cm data is $S_{\nu}\propto \nu^{(0.62\pm 0.02)}$, consistent with an optically thick
thermal source and very close to the canonical spectral index value as  expected from a stellar wind.

This result is quite different from what was claimed by Wright \& Barlow (\cite{Wri1975})
who derived a spectral slope of 0.96 from  observations of \object{$\beta$~Lyrae}  in the early seventies
carried out at 2.3 and 8.4~GHz.
We attribute the difference between this and our result to the large experimental errors associated with  the old measurements.

Jameson $\&$ King (\cite{Jameson}) made the first attempt to model the radio emission of \object{$\beta$~Lyrae}  by assuming a homogeneous, spherically 
symmetric \ion{H}{ii} region.
This  model, which assumes an emitting region embracing the entire system, foresees a black body spectral index and a  turnover frequency at 84~GHz, 
which is  not evident from our observations.

\subsection{Stellar wind/s}

The UV spectrum of \object{$\beta$~Lyrae} has been known to be dominated by 
anomalous continuum and very strong emission lines (Hack et al. \cite{Hack}).
The observed emission features, which show  unusually strong P-Cyg profiles, 
can be divided in two groups. The first group contains lines of highly ionized species, typical of hot-star winds, with a broad underlying emission.
The other group includes lines of species at lower ionization stages 
(Aydin et al. \cite{aydin}).

A  two-wind radiation-driven model has been proposed by Mazzali et al. (\cite{Mazza92})
to explain these two different kinds of  mass loss features. The different characteristics suggest  two  distinct  regions where they form, the first associated with the wind of the B0 V gainer, the second in the wind of the 
B6-8~II loser. 

The possibility that the radio emission could be related to the interactions  
between the two stellar winds is ruled out by the fact that the Mazzali et al. 
(\cite{Mazza92}) model outlines a system where only one of the winds is 
dominant. This would imply that the effects of possible wind interactions are 
negligible (Stevens \cite{steve}).
Moreover, the observed radio spectrum is consistent with thermal emission rather than non-thermal as observed in most colliding wind radio sources
(Williams \cite{Williams}).

Recently Umana et al. (\cite{Umana}) resolved an extended radio nebula,  embracing the entire binary system, by using the MERLIN interferometer at 4.9 GHz. In this paper the authors  suggested that it is possible to interpret the observed nebula, in the framework of the Mazzali et al. (\cite{Mazza92})
hypothesis, as a result of the  the stellar wind associated with the BII component ionized by the hotter companion.
Such a phenomenon is observed in Symbiotic systems where the associated radio nebula is the material ejected 
by one of the components which is  ionized by the secondary star.

The radio properties of symbiotics are well summarized in a series of papers 
(see for example, Seaquist \& Taylor \cite{Sea1990}). In particular,
the observed radio spectra have been modelled in terms of a binary model, where the morphology of the radio emitting region is a  function of the physical characteristics of the system such as 
binary separation ($a$), Lyman continuum luminosity of the hot component ($L_\mathrm{uv}$),  mass loss rate ($\dot{M}$) and velocity ($v$)
of the dominant wind (Taylor \& Seaquist \cite{Taylor1984} TS).

If the wind  is completely ionized, the ionized region shows the same radio properties as a spherically symmetric ionized wind.
Thus, in the framework of the TS model, a spectral index close to the canonical 0.6, as we observed in \object{$\beta$~Lyrae}, would imply that at least a large fraction of the wind is ionized.
A large ionization region can be  written in the TS terminology as $q_{o}a << a$, where  $q_{o}$ is the distance, measured in terms of the binary separation
$a$, from the center of the mass-losing star and the boundary of the ionized
region.
Using this assumption, from equation 8 of TS, we can relate the turnover 
frequency to the physical parameters of the system by the equation:
\begin{equation}
\nu_\mathrm{t}^{2.1} >> 8 \times 10^{66} 
(\frac{ T_\mathrm{e}^{-1.35} }{a ^{3} })(\frac{ \dot{M}}{v})^{2} 
\end{equation}
where $a$ is in cm, $\dot{M}$ in $M_{\sun} \mathrm{yr}^{-1}$, 
$v$ in km~s$^{-1}$.\\
Assuming {\bf $a= 58.5 R_{\sun}$ (Bisikalo et al.  \cite{Bisikalo}) }, the $\dot{M}$ and $v$ 
values of the dominant wind (BII component, see Table~\ref{nebula}) and 
$T_\mathrm{e}$ value obtained by MERLIN observations (Umana et al. 
\cite{Umana}), we derive a turnover frequency higher than 760 GHz.\\
The observed spectrum, which is optically thick up to $\sim 300$ GHz, is thus consistent with the symbiotic interpretation.

The far IR part of the spectrum is more difficult to interpret mostly because
at these frequencies the contribution of the photosperic emission becomes important.

The primary star in \object{$\beta$~Lyrae} is spectroscopically clearly visible and has been classified as a B8-6~II (Balachandran et al. {\cite{bala86}).
On the contrary very little is still known about the nature of the secondary 
nother than  the fact that it should be embedded in an accretion disk.
However, despite of the numerous efforts (Wilson \cite{Wilson74}; Linnell 
\& Hubeny \cite{Linnell1996}; Linnell et al. \cite{Linnell1998}; 
Linnell \cite{Linnell2000}), there is no standard model of an  accretion disk able 
to reproduce synthetic light curves that fit the observed ones from IR to UV.

\begin{figure}
\resizebox{\hsize}{!}{\includegraphics{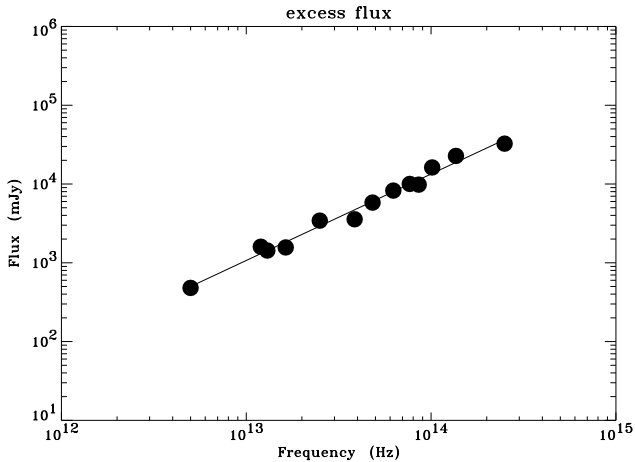}} 
\caption{The excess flux of  $\beta$~Lyrae. The  line is the fit
of the contribution at high frequencies of star+disk system. }
\label{excess}
\end{figure}

Therefore, we can compute the expected far-IR flux only for the primary 
component by assuming, between $1.5\times 10^{16}$ and $1.5\times 10^{12}$Hz 
a Kurucz (\cite{Kurucz93}) model with $T_\mathrm{eff}=13\,000~K$, $\log(g)=3$, 
as appropriate for the B8-6~II component. 

Since the observed continuum is the sum of the primary and secondary continuum, to normalize the atmospheric model we correct the observed V magnitude, as observed by Viotti et al. (\cite{Viotti78}), by a factor 1.10.
This has been calculated by assuming a flux ratio between the two components
at maximum light of 3 (Harmanec \cite{Harmanec90}) and a further decrement 
of about $12\%$ of the primary's flux due to the fact that the observations 
took place at orbital phase $\Phi=0.15$.

The radio emission from a stellar wind  comes from regions quite far from the star, where the wind velocity has reached its terminal velocity
($ v_\infty$). Because of the dependence of thermal opacity on wavelengths
($\kappa_{\lambda} \propto  \lambda^{2}$), emission at higher frequencies would come from regions much closer to the stellar photosphere, where some acceleration takes place.
Thus to evaluate the contribution of such a stellar wind to the far-IR part of the  spectrum, the effects of the velocity law of the velocity of the mass-flow must be taken into account.

 Lamers \& Waters (\cite{Lame1984}) have computed  the energy distribution for a grid of expanding, isothermal stellar wind models. 
In their work they considered all the possible ways the extended envelope can
affect the energy distribution of the central star, modelling quite accurately the far-IR part of the spectrum.

In adapting the  Lamers \& Waters (\cite{Lame1984}) model to the two winds
scenario of \object{$\beta$~Lyrae} we will assume that the secondary component
plays only  the  role of providing the radiation field necessary to ionize the 
wind associated with the primary. 
Following them, we thus derive the emission due to the
star+wind system, by assuming as physical parameters of the stellar wind those
of the dominant wind, that of the B8-6~II component, as derived by Umana et al. (\cite{Umana}).   
For the wind, we assume that the velocity is a function of  the radial distance,
in units of $R_{\ast}$:
\begin{equation}
\frac{v(x)}{v_\infty}=0.01+0.99(1-\frac{1}{x})
\end{equation}
as suggested by spectroscopic UV observations
(Castor \& Lamers \cite{Cala}). 

The obtained energy distribution, normalized at 5~GHz, is shown in 
Fig.~\ref{sed} (dotted line), where it is evident that the observed flux 
starts to deviate from the expected $\propto \nu^{2}$ photospheric behavior 
at $\sim 5~10^{14}$~Hz and there is an extra far-IR excess flux 
that the presence of a stellar wind cannot explain.\\
This excess, extracted and plotted in Fig.~\ref{excess}, can be fitted by a power law:
\begin{equation}
S_{\nu} \propto \nu^{1.10 \pm 0.04}
\end{equation}
in agreement, surprising, with the results from Viotti et al. (\cite{Viotti78}), which, however, were
based only on  JKLM observations  reported  in the plot.
As already pointed out by Viotti and co-authors, this result is consistent with a secondary hidden by an accretion disk, whose dimension increases with wavelength.

\begin{center}
\begin{table}
\caption{Binary and Nebula Characteristics}
\smallskip
\begin{tabular}{lcc}
{\em Binary\footnote{Mazzali et al. (\cite{Mazza92}}} & & \\
                       & {\em Gainer}            & {\em Loser}           \\ 
\hline
                       &                         &                       \\
$ T_\mathrm{eff}$[K]   & $32\,000$               & $13\,300$             \\
$ \dot{M}[M_{\odot}\mathrm{yr}^{-1}]$ 
                       &  $4.68 \times 10^{-8}$  & $7.16 \times 10^{-7}$ \\
$ v_{\infty}$[km s$^{-1}$]
                       & $1\,470$                & 388                   \\ 
                       &                         &                       \\
                       &                         &                       \\
 {\em   Radio Nebula\footnote{Umana et al. (\cite{Umana}}}   &     &                \\ 
\hline
                       &                         &                       \\

$ T_\mathrm{e}$[K]     & $11\,000 \pm 700$       &                       \\
$Size~(FWHM)$ [mas]	       & $(145 \pm 12) \times(100 \pm 8)$   & \\

\hline 
\end{tabular}
\label{nebula}
~\\
~\\
~\\
{\footnotesize 3) Mazzali et al. (\cite{Mazza92})}\\
{\footnotesize 4) Umana et al. (\cite{Umana})}\\
\end{table}
\end{center}

\subsection{Thermal jets}

The two-wind model of Mazzali has  been  recently criticized by Harmanec (2002),
who pointed out that there is no  evidence of wind associated with the 
secondary component, at least from emission or P Cygni lines as observed in the ultraviolet.  
Moreover, the elongated radio nebula, as observed by MERLIN, appears to be aligned with the reported optical jets, suggesting the possibility that
the radio emission is due to thermal radio jets emanating from the
accretion region around the hidden secondary.
 
 Several authors have investigated  the effects on radio flux, spectral 
index and mass-loss derivation due to deviations from sphericity in a 
extended stellar envelopes (Schmid-Burkg  \cite{schmid1982}; Reynolds \cite{reynolds1986}). 
While collimated  thermal jets can produce 
radio spectra significantly different from that of a canonical wind,
there are no spectral clues that can help in discriminating between a
spherically symmetric wind or biconical constant velocity flows
(wider thermal jets).
Therefore, we have a piece of observational evidence that if the 
origin of radio emission is a thermal jet it must be not collimated.\\
We can derive the jet opening angle ($\theta_{0}$) from the ratio of the minor
($\theta_{min}$) and major axis ($\theta_{max}$) of the radio nebula as derived
from MERLIN measurements (table 3).
\begin{equation}
\theta_{0}= 2 tg^{-1} \frac{ \theta_{min}}{\theta_{max}} \sim 69^{\circ}
\end{equation}
To determine the mass-loss of a thermal jet with opening angle $\theta_{0}$,
the formula, derived  in the hypothesis
of symmetry in the wind,  has to be corrected by a factor of $\eta=0.2 \, \theta_{0}
(sin i)^{-1/4}$ (eq. 21 of Reynolds, 1986), where $i$ is the angle of inclination of the jet as seen by the observer ($i \sim 85^{\circ}-88^{\circ}$) and $\theta_{0}$ is expressed in radians.\\
Therefore, we may derive the mass-loss from the relation:
\begin{equation}
\dot {M_{sphe}}= 6.7 \times 10^{-4} v_\infty 
S_\mathrm{6cm}^{3/4} D_\mathrm{kpc}^{3/2} (\nu\times g_\mathrm{ff})^{-0.5}  
~~~~ M_{\sun}\mathrm{yr}^{-1}
\end{equation}
where full ionization and cosmic abundances have been assumed,  $v_{\infty}$ is the terminal velocity of the wind, $S_\mathrm{6cm}$ the observed flux density 
at 6~cm in mJy, $D_\mathrm{kpc}$  the distance of the system, in  kpc. 
$g_\mathrm{ff}$ represents 
the free-free  Gaunt factor that, following Leitherer \& Robert (1991), can be    approximated  with:
\begin{equation}
g_\mathrm{ff}=9.77(1+0.13 \log{\frac{T^{3/2}}{\nu}})
\end{equation}
where $T$, in Kelvin, is the wind temperature.\\
By assuming a  typical stellar wind velocity ofa  hot B star  for the secondary component ($v\sim ~ 1500\mathrm{~km~s}^{-1}$) and constant temperature equal to the brightness temperature, derived by radio observations 
($T=T_\mathrm{B}$), we
obtain a mass loss rate of 
 \begin{equation}
\dot {M_{sphe}}= 2.0 \times 10^{-6}~~~~ M_{\sun}\mathrm{yr}^{-1}.
\end{equation}
which, corrected for thermal jets yields a mass-loss of
 \begin{equation}
\dot {M_{jet}}= \eta \dot {M_{sphe}}=5.2 \times 10^{-7}~~~~ M_{\sun}\mathrm{yr}^{-1}.
\end{equation}
A wide thermal jet will emit a radio flux whose frequency dependence is indistinguishable from that emerging from a spherically symmetric stellar wind.
Therfore, we can apply the model of
Lamers \&  Waters (1984), introducted in the previous section, to derive
the emission due to the star+jets system where,
this time, the physical parameter of the stellar wind associated with the gainer
(see table 3) and mass-loss corrected for thermal jets 
have been assumed.\\
The obtained energy distribution, normalized at 5~GHz, is showed in 
Fig.~\ref{sed2} (dotted-line), where the contribution of the primary 
(as indicated by the KURUCZ model) has been also taken  into account.
For a better comparison, the model, obtained in the case of symbiotic winds
(see section 4.1) is shown in the figure as a dashed line.
The model  stops at $\nu \sim 10^{14} \rm{Hz}$ as this is the highest frequency
of the  Lamers \&  Waters (1984) tabulation.

As already noted for the symbiotic wind hypothesis, there is an extra far-IR excess flux that a thermal jet, associated with the secondary, is not able to explain. The extracted excess flux is well fitted by a power-law, with a spectral index $\alpha =0.91 \pm 0.06$, still consistent with the contribution of an accretion disk, whose density decreases with radius.
 
\begin{figure*}
\resizebox{\hsize}{!}{\includegraphics{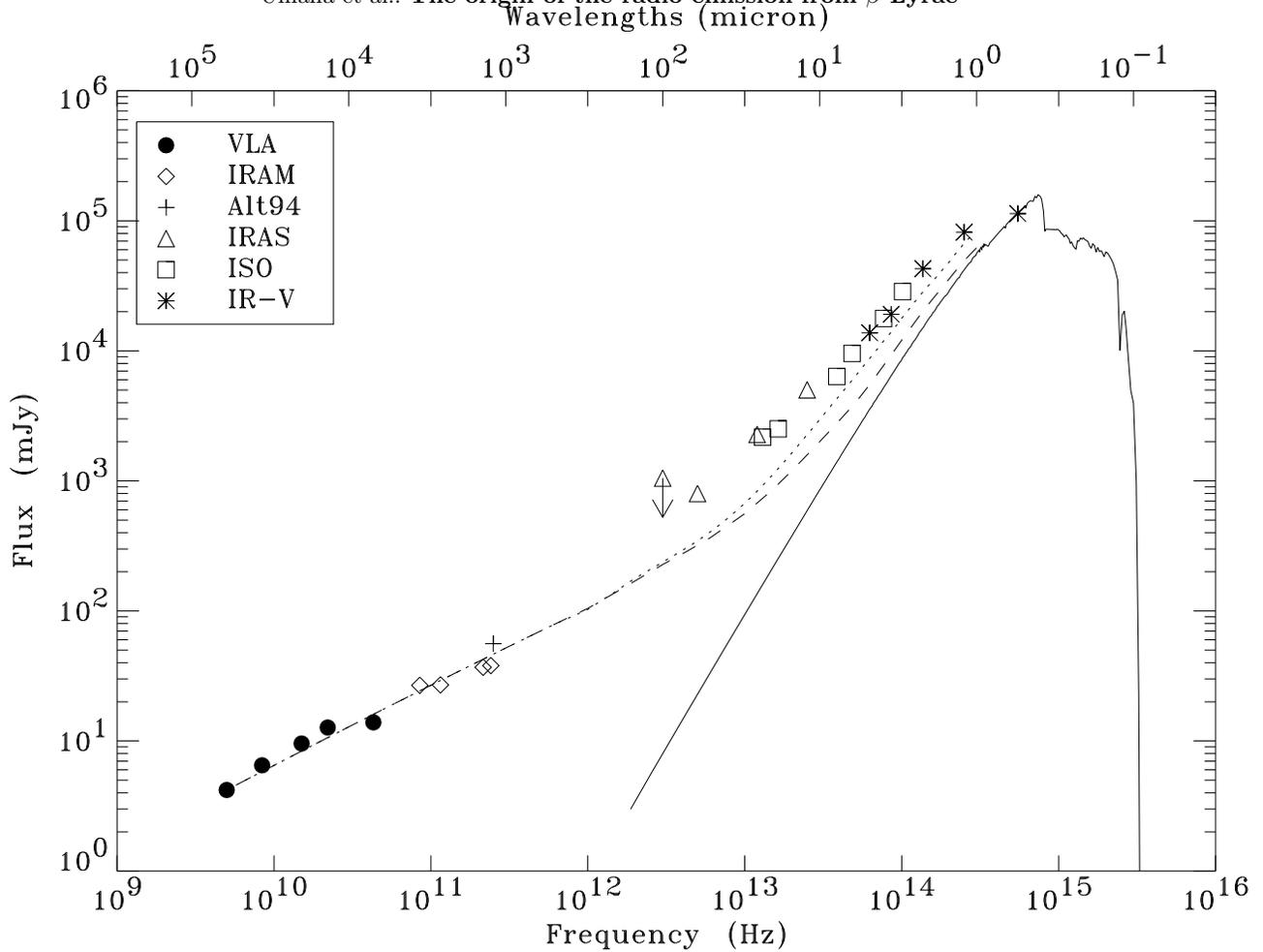}} 
\caption{
Same as Fig. ~\ref{sed}.
The dotted line is the contribution at high frequencies of the free-free plus bound-free emission from a biconical outflow plus the secondary  contribution
represented 
by an ATLAS9 (Kurucz \cite{Kurucz93}) model (continuos line), with  fluxes normalized to the flux in the V band (solid line). For comparison the fit 
in the case of symbiotic wind is also shown (dashed line)}
\label{sed2}
\end{figure*}

\section{Conclusions}

\object{$\beta$~Lyrae} is one of the best studied stellar systems: still no 
reliable model of the system, able to reproduce the huge amount of 
observational data, is available.\\
  Very recently, a new generation of models, that take into consideration the details of the physics of the accretions disk,  
have been presented (Linnell et al. \cite{Linnell1998}; Bisikalo et al. \cite{Bisikalo}). Even if they provide a good model for the light curves in 
the near and far-UV and in the visible, they still do not predict the shape of 
the infrared light curve, and, in particular, they do not reproduce the secondary 
minimum that becomes deeper than the primary minimum at $\lambda \geq 4.8 \mu m$.
This leads to the conclusion that another component, besides stars plus accretion disk, must contribute to the observed flux, and Linnell 
(\cite{Linnell2000}) proposes that the extra source of continuum radiation may 
be Thomson scattering of radiation from the gainer.

In this paper we have presented new observational evidence that support the presence of an extra-component in the already quite complex binary system.
This extra component, which is the  origin of  the observed thermal radio flux, can be due to an extended stellar wind ionized by the strong ultraviolet flux of the secondary component or related to collimated structures associated with the 
gainer (conical thermal jets).

We further evaluated the contribution of this component  in the far-IR to extract the 
spectral energy distribution of the stars plus accretion disk system. Our 
results, in both hypotheses,  show  a power-law distribution up to IRAS frequencies indicating 
an accretion disk with a non-uniform density distribution and whose size 
varies with wavelength. 
To quantify this, the collection of new, good quality infrared light curves would be highly valuable.

The stellar wind or thermal jets,  as appearing from the obtained spectrum and its modelling, is optically thick up to IRAS-ISO frequencies. Still, at those frequencies we measure a significant excess that we attribute to the disk since it is  in agreement with the near-IR data.
We may ask what kind of effect  the radio nebula may have  on the observability 
of the inner disk. 
In both the cases, the size of the radio source depends on  frequency as $R_\mathrm{radio} \propto \nu^{-0.7}$. Thus, scaling the size 
measured at 4.8 GHz with MERLIN (see Table~\ref{nebula}) we obtain that the 
radio nebula reaches dimensions comparable to those of the entire binary system
at frequencies of the order of $10^{13}$~Hz, close to the IRAS spectral region.
  
This provides important constraints to the radio source associated with the 
binary system,   whose morphology should allow one to observe also the disk.
The other possibility, that the disk is surrounding the nebula, 
can be discarded, since a clear eclipse is observed in the infrared.

Finally, we would like to point out that even in the case of stellar wind, 
the fact that the radiation field necessary to ionize the wind is 
provided by the  secondary component, which is embedded in the thick accretion disk  (Linnell \cite{Linnell2000}), suggests that the ionization would take place mainly in the polar regions. 
Thus, in both the cases,  it is quite probable that the nebula has a 
bipolar morphology that the MERLIN observations (Umana et al. \cite{Umana}) 
were not able to resolve.
This bipolar morphology, which  should have a wide opening angle as suggested 
by the observed spectral index (Rodriguez et al. \cite{Rod1990}), is compatible 
with the possibility of observing the inner accretion disk.

\section{Acknowledgements}
We acknowledge the IRAM staff from Plateau de Bure for carrying out the observations and for help provided during the data reduction.
This article used archive observations with ISO, an ESA project with instruments funded by ESA Member States (especially the PI countries: France, Germany, the Netherlands and the United Kingdom) and with the participation of
ISAS and NASA. The ISO Spectral Analysis Package (ISAP) is a joint development by the LWS and SWS Instrument Teams and Data Centers. Contributing institutes are CESR, IAS, IPAC, MPE, RAL and SRON.
We wish to thank the referee Dr. Petr Harmanec as his suggestions greatly 
helped us in improving the paper.

\smallskip


\end{document}